\documentstyle[11pt]{article}

\renewcommand{\theequation}{\thesection.\arabic{equation}}

\newlength{\extraspace}
\newlength{\extraspaces}
\newcounter{dummy}

\newcommand{\baa}{
\addtocounter{equation}{1}
\setcounter{dummy}{\value{equation}}
\setcounter{equation}{0}
\renewcommand{\theequation}{\thesection.\arabic{dummy}\alph{equation}}
\begin{eqnarray}
\addtolength{\abovedisplayskip}{\extraspaces}
\addtolength{\belowdisplayskip}{\extraspaces}
\addtolength{\abovedisplayshortskip}{\extraspace}
\addtolength{\belowdisplayshortskip}{\extraspace}}

\newcommand{\eaa}{
\end{eqnarray}
\setcounter{equation}{\value{dummy}}
\renewcommand{\theequation}{\thesection.\arabic{equation}}}

\newcommand{\be}{\begin{equation}
\addtolength{\abovedisplayskip}{\extraspaces}
\addtolength{\belowdisplayskip}{\extraspaces}
\addtolength{\abovedisplayshortskip}{\extraspace}
\addtolength{\belowdisplayshortskip}{\extraspace}}
\newcommand{\ee}{\end{equation}}

\newcommand{\ba}{\begin{eqnarray}
\addtolength{\abovedisplayskip}{\extraspaces}
\addtolength{\belowdisplayskip}{\extraspaces}
\addtolength{\abovedisplayshortskip}{\extraspace}
\addtolength{\belowdisplayshortskip}{\extraspace}}
\newcommand{\ea}{\end{eqnarray}}

\newcommand{\bd}{\begin{displaymath}
\addtolength{\abovedisplayskip}{\extraspaces}
\addtolength{\belowdisplayskip}{\extraspaces}
\addtolength{\abovedisplayshortskip}{\extraspace}
\addtolength{\belowdisplayshortskip}{\extraspace}}
\newcommand{\ed}{\end{displaymath}}

\newcommand{\ban}{\begin{eqnarray*}
\addtolength{\abovedisplayskip}{\extraspaces}
\addtolength{\belowdisplayskip}{\extraspaces}
\addtolength{\abovedisplayshortskip}{\extraspace}
\addtolength{\belowdisplayshortskip}{\extraspace}}
\newcommand{\ean}{\end{eqnarray*}}

\newcommand{\newsection}[1]{
\vspace{15mm}
\pagebreak[3]
\addtocounter{section}{1}
\setcounter{equation}{0}
\setcounter{subsection}{0}
\setcounter{footnote}{0}
\begin{center}
{\Large \thesection. #1}
\end{center}
\nopagebreak
\medskip
\nopagebreak}

\newcommand{\deel}[2]{{\textstyle{#1 \over #2}}}

\newcommand{\ie}{{\it i.e.}}
\newcommand{\eg}{{\it e.g.\ }}
\newcommand{\re}{\mbox{I}\!\mbox{R}}
\newcommand{\nne}{\mbox{I}\!\mbox{N}}

\def\inbar{\,\vrule height1.5ex width.4pt depth0pt}
\font\rms=cmr12 at 12pt
\def\ce{\relax\ifmmode\mathchoice
{\hbox{$\inbar\kern-.3em{\rm C}$}}
{\hbox{$\inbar\kern-.3em{\rm C}$}}
{\lower.9pt\hbox{\rms $\inbar\kern-.3em{\rm C}$}}
{\lower1.2pt\hbox{\rms $\inbar\kern-.3em{\rm C}$}}
\else{$\inbar\kern-.3em{\rm C}$}\fi}
\font\cmss=cmss12 \font\cmsss=cmss12 at 12pt
\def\ze{\relax\ifmmode\mathchoice
{\hbox{\cmss Z\kern-.4em Z}}{\hbox{\cmss Z\kern-.4em Z}}
{\lower.9pt\hbox{\cmsss Z\kern-.4em Z}}
{\lower1.2pt\hbox{\cmsss Z\kern-.4em Z}}\else{\cmss Z\kern-.4em Z}\fi}

\newcommand{\dif}{\partial}
\newcommand{\dbar}{\bar{\dif}}

\newcommand{\hb}{\bar{h}}
\newcommand{\gb}{\bar{g}}

\newcommand{\gam}{\Gamma}

\newcommand{\tr}{\mbox{Tr}}
\newcommand{\del}{\delta}

\newcommand{\mub}{\bar{\mu}}

\newcommand{\actie}[1]{\deel{1}{2\pi}\int d^2z \, }

\newcommand{\vars}[2]{{\del {#1} \over \del {#2}}}

\newcommand{\np}[1]{Nucl. Phys. {\bf B#1}}
\newcommand{\cmp}[1]{Comm. Math. Phys. {\bf #1}}
\newcommand{\intmod}[1]{Int. Journal of Mod. Phys. {\bf A#1}}
\newcommand{\plb}[1]{Phys. Lett. {\bf B#1}}

\newcommand{\ad}[1]{\mbox{\rm ad}_{#1}}

\begin{document}

\addtolength{\baselineskip}{.7mm}

\thispagestyle{empty}
\begin{flushright}
{\sc THU}-92/26\\
{\sc ITFA}-92/24\\
9/92
\end{flushright}
\vspace{1cm}
\setcounter{footnote}{2}
\begin{center}
{\LARGE\sc{$W$ algebras, $W$ Gravities and Their Moduli Spaces}}\\[1.5cm]

\sc{F. A. Bais, T. Tjin\footnote{e-mail: tjin@phys.uva.nl}, P. van
Driel\footnote{
Address from September 1, 1992: Physics Dept. UCLA, University of
California, Los Angeles, CA 90024, USA}}\\[4mm]
{\it Instituut voor Theoretische Fysica\\[1.5mm]
     Valckeniersstraat 65, 1018 XE Amsterdam\\[1.5mm]
     the Netherlands}\\[4mm]
\sc{Jan de Boer\footnote{e-mail: deboer@ruunts.fys.ruu.nl} and
Jacob Goeree\footnote{e-mail: goeree@ruunts.fys.ruu.nl}}
\\[4mm]
{\it Institute for Theoretical Physics, University of Utrecht\\[1.5mm]
Princetonplein 5, P.O. Box 80.006\\[1.5mm]
3508 TA Utrecht, the Netherlands}\\[15mm]

{\sc Abstract}\\[1cm]
\end{center}

\noindent
By generalizing the Drinfel'd--Sokolov reduction we construct a large class of
$W$ algebras as reductions of Kac-Moody algebras. Furthermore we construct
actions, invariant under local left and right $W$
transformations, which are the classical covariant
induced actions for $W$ gravity.\\[20mm]

\noindent
Talk presented by T. Tjin at the `Workshop on Superstrings and Related
Topics', Trieste, July 1992.

\vfill

\newsection{Introduction}

$W$ algebras were introduced by Zamolodchikov \cite{Za} as extensions of
the Virasoro algebra which plays a central role in Conformal Field theory.
Since $W$ algebras are extensions of the Virasoro algebra a Hilbert space
which is a direct sum of infinitely many irreducible Virasoro
representations may be just a finite direct sum of irreducible $W$
representations. Ever since the original paper by Zamolodchikov
there has been an outpouring of papers on the subject (for a recent
review see \cite{BoSc}). It is now clear that these $W$ algebras
play a central role in many areas of two-dimensional physics, most notably
Toda theories \cite{toda}, gauged WZW models \cite{wzw}, reductions
of the KP hierarchy \cite{kp} and in the matrix model formulation of
two-dimensional quantum gravity \cite{ijdel}.

Many $W$-algebras were constructed by
what has become known as the `direct method'
which amounts to solving the
Jacobi identities for an algebra in which part of the commutation
relations are fixed by imposing that all generators are primary fields.
Although this method does give a lot of explicit information, it is extremely
cumbersome. Therefore, different methods were developed which
construct $W$-algebras as reductions of others. One of those methods is
the so called `coset construction', see \cite{GKO,BBSS}. Another method,
called quantum Drinfel'd--Sokolov reduction \cite{DS,FF,FL,beroog},
constructs the aforementioned
`standard' $W$ algebras as Hamiltonian reductions of Kac--Moody
algebras. This method has the important advantage
that, due to general theorems of Hamiltonian reduction \cite{AM},
closure of the algebra and Jacobi identities are ensured.

In \cite{poldiff,bersh} it was shown that apart from the Drinfel'd--Sokolov
reduction of the $SL_3(\re)$ Kac--Moody algebra, leading to the `standard'
$W_3$ algebra, there exists another
reduction leading to an algebra closely resembling (a bosonic version of
the) $N=2$ super
conformal algebra. This algebra was named $W_3^{(2)}$. Shortly after,
it was shown that one can associate a reduction to every
embedding of $sl_2$ into the finite Lie algebra underlying the Kac--Moody
algebra \cite{adam,dublin}. These reductions generalize the
Drinfel'd--Sokolov reductions, and contain among others the aforementioned
ones.

Given this large set of non-linear $W$ algebras it has been
an open problem what their geometrical interpretation is. Whereas we know
that the Virasoro algebra arises after gauge fixing the two-dimensional
diffeomorphism invariance, a similar mechanism for the $W$ algebras is
unknown. Many groups attempted to get a better understanding
of $W$ geometry by the construction of an action invariant
under {\em local}\, left and right $W$ transformations. Hull \cite{hul}
was the first to construct an action invariant under {\em chiral}
$W_3$ transformations. Schoutens {\em et al.} extended these results
to the non-chiral case \cite{alles}, and also discussed some of the
quantum properties of these theories.

In \cite{jj3} a different point of view was taken and
it was shown that the non-linear $W$ transformations are
just homotopy contractions of ordinary gauge tranformations.
This result was used to construct a left, right invariant action
for arbitrary $W$ algebras, whose properties show it can be seen as the
covariant induced action for $W$ gravity.
It was shown
that the action is simply a Legendre transform of the WZW action based on
the group underlying the Kac--Moody algebra. The moduli space for $W$
gravity was shown to be the quotient of a subspace of the space of flat
$SL_n(\ce)$ connections by the modular group. For $n=2$ this is the usual
moduli space of Riemann surfaces.

More details on the work described here can be found in
\cite{adam,jj3}.

\newsection{Generalized Drinfel'd--Sokolov Reductions}

A classical $W$-algebra with conformal weights $\{\Delta_i \}_{i
\in I} \subset \frac{1}{2} \nne$ is a Poisson algebra
generated by fields $T(z), \{A_i(z)\}_{i \in I}$ such that
\begin{enumerate}
\item $\{T(z),T(w)\}=T'(w) \delta (z-w) +2T(w) \delta '(z-w)+\deel{c}{12}
\delta '''(z-w)$
\item $\{T(z),A_i(w)\}=\Delta_i A_i(w) \delta'(z-w)+A'_i(w)
\delta (z-w)$
\item Jacobi identities are satisfied.
\end{enumerate}
Finding relations $\{A_i(z),A_j(w)\}$ such that part 3 of this
defintion is satisfied is in general rather complicated. We
shall therefore construct $W$-algebras as Poisson reductions of
the Kirillov Poisson algebra on an $sl_n$ Kac Moody algebra.
This algebra is generated by fields $\{J^a(z)\}$ satisfying
\be
\{J^a(z),J^b(w)\}=f^{ab}_cJ^c(w) \delta (z-w) -kg^{ab}\delta
'(z-w) \label{KM}
\ee
where $g^{ab}$ is the inverse of the matrix $g_{ab}=Tr(t_at_b)$,
$\{t_a\}$ is a basis of $sl_n$
and
$f_{ab}^c$ are the structure constants in this basis.
This algebra, which is well known to be the current algebra of
the $SL_n$ WZW model,
is itself a $W$-algebra with
$T(z)=g_{ab}J^a(z)J^b(z)$ the so called Sugawara stress energy
tensor.

Let $i: sl_2 \hookrightarrow sl_n $ be an $sl_2$ embedding into
$sl_n$. Under the adjoint action of $sl_2$ the algebra $sl_n$
(seen as
a representation of $sl_2$) branches into a direct sum of $p$
irreducible $sl_2$ multiplets. Let $\{t_{k,m}\}_{m=-j_k}^{j_k}$
be a basis of the $k^{th}$ multiplet where $j_k$ is the highest
weight of this multiplet. The numbering is chosen such that
$t_{1,\pm 1}=t_{\pm}$ and $t_{1,0}=t_3$ where $\{t_3,t_{\pm}\}$
is $i(sl_2)$. An arbitrary map $J:S^1 \rightarrow sl_n$ can then
be written as
\begin{equation}
J(z)=\sum_{k=1}^{p}\sum_{m=-j_k}^{j_k} J^{k,m}(z)t_{k,m
}\label{current}
\end{equation}
Impose now the constraints
$\phi^{1,1}(z) \equiv J^{1,1}(z)-1=0$ and $\phi^{k,m}(z) \equiv
J^{k,m}(z)=0$ for $m>0,k \neq 1$. The constraints
$\{\phi^{k,m}(z)\}_{m \geq 1}$ turn out to be first class which
means that they generate gauge invariance. This gauge invariance
can be completely fixed by gauging away the fields
$\{J^{k,m}(z)\}_{m >-j_k}$. After constraining and gauge fixing the
currents look like
\begin{equation}
J_{fix}(z)=\sum_{k=1}^{p}J^{k,-j_k}(z)t_{k,-j_k}+t_+
\label{fix}
\end{equation}
The Poisson bracket (\ref{KM}) on the set of elements
(\ref{current}) induces a Poisson bracket $\{.,.\}^*$ on the set
of elements (\ref{fix}) which is nothing but the Dirac bracket.
One now has to calculate the Dirac brackets between the fields
$\{J^{k,-j_k}\}$. It is one of our main results that the fields
$T(z)=\mbox{Tr}(J^2_{fix}(z))$ and $\{J^{k,-j_k}(z)\}_{k>1}$ generate
a $W$ algebra with conformal weights $\{\Delta_k=j_k+1\}_{k>1}$ w.r.t.
the Dirac bracket.

\newsection{A General Expression for $W$ Algebras}

It is possible to give a closed expression for the Poisson
bracket of a general $W$ algebra. For this we need a
linear operator $L: sl_n\rightarrow sl_n$ which is the inverse
of $\ad{t_+}$. In terms of the basis
$\{t_{k,m}\}^{j_k}_{m=-j_k}$ of $sl_n$ it is defined by
requiring (i) $L(\ad{t_+}(t_{k,m}))=t_{k,m}$ for $m<j_k$ and
(ii) $L(t_{k,-j_k})=0$. Let us also define a field $W(z)$ by
\be
W(z)=J_{fix}(z)-t_+=\sum_{k=1}^{p} J^{k,-j_k}t_{k,-j_k}.
\ee
The Poisson bracket of two functionials $Q_1$ and $Q_2$ of the
fields $\{J^{k,-j_k}(z)\}_{k>1}$ is given by
\be \label{explicit}
\{Q_1,Q_2\}^{\ast}=\int dz\tr\left(\vars{Q_2}{W(z)}
(k\dif+\ad{W(z)})\frac{1}{1+L(k\dif+\ad{W(z)})}
\vars{Q_1}{W(z)} \right),
\ee
where
\be
\vars{Q}{W(z)}=\sum_{k=1}^{p} \vars{Q}{J^{k,-j_k}(z)} t^{k,-j_k}
\ee
and $\{t^{k,m}\}$ is the dual basis of $\{t_{k,m}\}$, so that
$\tr(t^{k,m}t_{k',m'})=\delta^{k}_{k'}\delta^{m}_{m'}$. In
(\ref{explicit}), $1/(1+L(\dif+\ad{W(z)}))$ denotes the series
$\sum_{i\geq 0} (-L(\dif+\ad{W(z)})^i $. This series always
truncates after a finite number of steps: $(-L(\dif+\ad{W(z)}))^i=0$
for $i\geq 2\Delta_{max}$, where $\Delta_{max}$ is the largest
conformal weight of the $W$ algebra under consideration.

\newsection{Classical $W$ Gravity}

Introduce a new constrained current $\overline{J}_{fix}$
of the form $\overline{J}_{fix}(z)=
\sum_{k=1}^{p}\overline{J}^{k,-j_k}(z)
\overline{t}_{k,-j_k}+\overline{t}_{+}$, where $\overline{t}_{k,m}$
is the transpose of $t_{k,m}$.
For this constrained current one can write down a Poisson
bracket analogous to (\ref{explicit}), showing that the $\{
\overline{J}^{k,-j_k} \}$ generate a $\overline{W}$
algebra isomorphic to the $W$ algebra generated by $J_{fix}(z)$.
One might wonder whether the $W,\overline{W}$
transformations as following from the previously discussed Poisson
brackets, \eg\ (\ref{explicit}), have an interpretation as
symmetries of a field theory. Such a theory would then
commonly be called `a theory for $W$ gravity'.
It can be shown that such
an interpretation of the above algebras indeed exists: an action
invariant under these $W,\overline{W}$ transformations
can be constructed.

Consider the following definition of the {\em chiral}\, induced
action $\gam[\mu_i]$
for $W$ gravity:
\be \label{chiral}
e^{-\gam[\mu_i]}=\left< e^{-\int d^2z \sum_i \mu_i W_i}
\right>.
\ee
In this formula the $\mu_i$ are generalizations of the Beltrami
differential $\mu=\mu_2$, and the $W_i$ are matter currents forming
a $W$ algebra. The induced action can be calculated by expanding
the exponent and making use of the operator product algebra of the
$W_i$ fields, which in the classical limit we consider here
reduces to the Poisson algebra as given in (\ref{explicit}).
Doing so, one derives classical Ward identities for the induced action
whose solution is given by: $\gam[\mu_i]=
kS_{wzw}(hf)$. Here $S_{wzw}$ is the well known WZW action,
$f=\exp(-zt_+)$, and $h\in SL_n(\re)$ is such that
\be \label{h}
h^{-1}\dbar h=\frac{1}{1+L(k\dif+\ad{W(z)})}F(\mu_i),
\ee
where $F(\mu_i)\in \mbox{ker}(t_+)$, whose independent
components are labeled by the $\mu_i$.

The chiral action is {\em not}\, invariant under the $W$ transformations,
but rather has an anomalous transformation behaviour. The precise form
of this anomaly is directly related to the anomalous terms in the Poisson
brackets of the $W$ fields, \eg\ the $c/12\, \delta'''(z-w)$ term of the
Virasoro algebra.

In a similar way we can construct a chiral action $\gam[\mub_i]$ starting
from matter currents $\overline{W}_i$ which form a $\overline{W}$ algebra.
To obtain from these two chiral actions an action which is invariant under
$W,\overline{W}$ transformations, we should construct a local counterterm
$\Delta \gam$ whose anomalous behaviour under $W,\overline{W}$ transformations
cancels that of the chiral actions. So schematically
\be \label{schema}
S=\gam[\mu_i]+\Delta \gam+\gam[\mub_i].
\ee
This counterterm can be constructed as follows:
Let $g,\gb$ be elements of $SL_n(\re)$ such that $g^{-1}\dif g=J_{fix}$ and
$\gb^{-1}\dbar \gb=\overline{J}_{fix}$, and let
$\bar{h}$ be the `conjugate' (by which we mean taking the
transpose and putting bars) of $h$ as defined in (\ref{h}).
Then the final result for the covariant induced action of arbitrary $W$
gravity reads
\be \label{action}
S(\mu_i,\mub_i,G)=-k\,\mbox{min}_{W_i,\overline{W}_i}
\left( S_{wzw}(gG\gb^{-1})-S_{wzw}(gh^{-1})-S_{wzw}(\gb^{-1}\hb)
\right),
\ee
\ie\ it is the Legendre transform w.r.t. the fields $W_i,\overline{W}_i$
of the WZW action. Recall that $W_i,\overline{W}_i$ label the
independent components of $J_{fix}$ and $\overline{J}_{fix}$,
respectively, and are dual to the parameters $\mu_i,\mub_i$.
In (\ref{action}) $G$ is an extra group element of $SL_n(\re)$
needed to make the action invariant.
One can work out the Legendre transform
quite easily since the $W_i,\overline{W}_i$ appear in an
algebraic and at most quadratic way
in (\ref{action}). Specializing to a particular $W$ algebra
one finds in general that some of the components of $G$ appear
only algebraically, making it possible to integrate them out.
For instance if one considers the case of $SL_2(\re)$, for which the
corresponding $W$ algebra is simply the Virasoro algebra,
(\ref{action}) reduces to the well known Polyakov action
for two-dimensional gravity \cite{pol1}
\be \label{pol}
S_{pol}\sim \int d^2z\, R\frac{1}{\Box}R,
\ee
once we intergrate out the auxiliary fields.

\newsection{Moduli Spaces for $W$ Algebras}

In the same way as the moduli space of Riemann surfaces plays an
important role for conformal field theories coupled to gravity,
on expects a generalized moduli space to play an important role
for conformal field theories coupled to $W$ gravity. Such a
generalized $W$ moduli space can be defined as the space of
$W$ fields on a Riemann surface modulo $W$ transformations.
Let us sketch some of the steps
involved in the computation.

First, one has to construct $W$ algebras on arbitrary Riemann
surfaces. Because currents transform as a connection one-form
with values in
$sl_n$, the constraint $J^{1,1}=1$ cannot be imposed globally.
The resolution of this problem is to twist the trivial $sl_n$
bundle for which $J$ is a connection
into a non-trivial vector bundle $E$,
such that the $(J-J_{ref})^{1,1}-1=0$ can be imposed globally.
The fixed reference connection $J_{ref}$ is needed, because $0$
is no longer a globally well-defined connection for $E$. It
turns out that $E$ is isomorphic to the direct sum of line
bundles $\oplus_k \oplus_{m=-j_k}^{j_k} K^m$, where $K$ is the
holomorphic cotangent bundle of the Riemann surface. Using $E$,
one can define $W$ algebras on arbitrary Riemann surfaces using
a generalization of (\ref{explicit}).

Next, one shows that the space of $W$ fields modulo $W$
transformations is the same as the space of holomorphic $W$
fields modulo holomorphic $W$ transformations. Because these
spaces are finite dimensional, they are easy to investigate.
The $W$ moduli spaces computed in this way
are actually only $W$
Teichm\"uller spaces; to obtain $W$ moduli space one has to take
the quotient by the action of the modular group.
To prove that no moduli are lost when passing to
holomorphic $W$ fields, a geometrical construction of $W$
algebras is used: $W$ algebras can be obtained as a homotopy
contraction (see e.g. \cite{botttu})
of ordinary gauge transformations. The homotopy
operator is precisely the operator $L$ defined in the previous
section, as it can be used to defined a map from the space of
$(1,0)$ forms with values in $E$ to the space of sections of
$E$.

For genus $g>1$, the $W$ moduli spaces are $(g-1)(n^2-1)$
dimensional subspaces of the $2(g-1)(n^2-1)$ dimensional
space of irreducible flat $SL_n(\ce)$ over a Riemann surface.
Furthermore, they are related in a natural way to so-called
Higgs bundles \cite{hitchin,hitchin2,hitchin3,simpson,simpson2}.
This correspondence can be used to show that for
the principal $sl_2$ embeddings in $sl_n$, the $W$ moduli space
is the moduli space of flat $SL_n(\re)$ bundles. For other
embeddings, the interpretation of the $W$ moduli spaces is
unclear.

\end{document}